\documentclass[aps,twocolumn,prb,superscript,floatfix,superscriptaddress,10pt]{revtex4-2}

\usepackage{epsfig}
\usepackage{graphicx}
\usepackage{dcolumn}
\usepackage{bm}
\usepackage{amsfonts}
\usepackage{amsmath}
\usepackage{amssymb}
\usepackage{mathrsfs}
\usepackage{natbib}
\usepackage{multirow}
\usepackage[T1]{fontenc} 
\usepackage{lmodern}
\usepackage{xcolor}
\usepackage{hyperref}
\usepackage{pifont}
\newcommand{\cmark}{\ding{51}}
\newcommand{\xmark}{\ding{55}}
\hypersetup{colorlinks=true,citecolor=blue,linkcolor=blue,urlcolor=blue}

\makeatletter
\AtBeginDocument{%
  \@ifpackageloaded{array}{%
    \let\@array@sw\@array@sw@array
    \let\@array\@array@array
    \let\@@array\@array
    \let\@tabular\@tabular@array
    \let\@tabarray\@tabarray@array
    \let\array\array@array
    \let\endarray\endarray@array
    \let\endtabular\endtabular@array
    \let\@mkpream\@mkpream@array
    \let\@classx\@classx@array
    \let\insert@column\insert@column@array
    \let\@arraycr\@arraycr@array
    \let\@xarraycr\@xarraycr@array
    \let\@xargarraycr\@xargarraycr@array
    \let\@yargarraycr\@yargarraycr@array
    \expandafter\let\csname endtabular*\endcsname\endtabular
  }{}%
}
\makeatother

\DeclareMathOperator{\Tr}{Tr}      

\newcommand{\vb}[1]{\bm{#1}}                     
\newcommand{\pdv}[2]{\frac{\partial #1}{\partial #2}} 
\newcommand{\eval}{\bigg|}                       

\let\phi=\varphi
\let\epsilon=\varepsilon

\DeclareMathOperator{\re}{Re}
\DeclareMathOperator{\im}{Im}

\newcommand{\vf}{v_{\textrm{F}}}
\newcommand{\vh}{v_{\textrm{H}}}
\newcommand{\vnh}{v_{\textrm{NH}}}

\begin{document}

\title{Minimal Hamiltonian deformations as bulk probes of effective non-Hermiticity in Dirac materials}

\author{Sergio Pino-Alarc\'on}
\thanks{These authors contributed equally to this work.}
\affiliation{Departamento de F\'isica, Universidad T\'ecnica Federico Santa Mar\'ia, Casilla 110, Valpara\'iso, Chile}

\author{Juan Pablo Esparza}
\thanks{These authors contributed equally to this work.}
\affiliation{Departamento de F\'isica, Universidad T\'ecnica Federico Santa Mar\'ia, Casilla 110, Valpara\'iso, Chile}
\affiliation{Instituto de F\'isica, Pontificia Universidad Cat\'olica de Valpara\'iso, Avenida Universidad 331, Curauma, Valpara\'iso, Chile}

\author{Vladimir Juri\v{c}i\'c}
\thanks{Contact author: vladimir.juricic@usm.cl}
\affiliation{Departamento de F\'isica, Universidad T\'ecnica Federico Santa Mar\'ia, Casilla 110, Valpara\'iso, Chile}

\date{Received 4 February 2026; revised 19 March 2026; accepted 27 April 2026; published 15 May 2026}

\begin{abstract}
Non-Hermitian (NH) Dirac semimetals describe open gain--loss systems. Yet at charge neutrality, models featuring real spectrum often look
Hermitian-like, with NH effects absorbed into renormalized band parameters. Here, we show that a response-based diagnostic of effective non-Hermiticity can be formulated using minimal
pseudo-Lorentz-symmetry-breaking deformations, which separate observables that remain captured by parameter
redefinitions from those that exhibit irreducible NH structure. For a two-dimensional NH Dirac semimetal in the weak-NH, real-spectrum regime, we analyze Dirac-cone tilt and velocity anisotropy and compute representative probes of spectral structure, quantum geometry, optical response, and viscoelasticity at zero temperature. We find that tilt yields an NH-dependent slope of the density of states that cannot be collapsed to a single effective velocity, while velocity anisotropy can be captured by
effective-velocity reparametrization. Furthermore, the quantum metric and collisionless optical conductivities provide
NH-insensitive benchmarks (with the nonlinear conductivity symmetry selected), whereas the shear viscosity offers a discriminator through its tensor structure. Our results identify minimal deformations and bulk response channels that
enable access to effective non-Hermiticity even when the spectrum remains real.

\end{abstract}

\maketitle

\section{Introduction}

Non-Hermitian (NH) extensions of quantum many-body systems provide an effective framework for
capturing coherent dynamics in the presence of loss, gain, and more general environmental couplings in
condensed-matter and synthetic platforms~\cite{ElGanainy2018NatPhys,Ashida2021AdvPhys,Bergholtz2021RMP,Miri2019Science}. { Recent work has also broadened the scope of NH physics across a wide range of settings, including Floquet and aperiodic systems, as well as disorder-driven and localization phenomena~\cite{fleury2023,Ji2025PRB,Tong2025PRB,Peng2025PRB,PhysRevB.111.094109,Chen2025PRB,Longhi2025PRL,Ou2025PRL,Zheng2025PRB,Shen2025PRB,Hu2025CommunPhys,LeongRoy2025,Cheng2026PRB,PRB112075123,PRB112075115}.}
NH Dirac materials~\cite{jurivcic2024yukawa,Murshed2024Quantum,Yu2024NHStrongly,Murshed2025Yukawa,leong2025arxiv,Roy-PRD-2025,Esparza2025EMAs}
are a particularly attractive arena for these ideas, given the broad realizations of Dirac
quasiparticles in electronic, photonic, and cold-atom settings~\cite{CastroNeto2009RMP,Armitage2018RMP,Polini2013NatNano,Tarruell2012Nature,Plotnik2014NatMater}. Furthermore, the hallmark non-Hermitian skin effect (NHSE), the extensive boundary accumulation driven by
nonreciprocity and point-gap topology~\cite{Martinez-Alvarez2018NHRobust,Yao2018Edge,Lee2019Anatomy,Borgnia2020NHBoundary,Zhang2020Correspondence,Okuma2020Topological}, is often strongly suppressed in NH Dirac settings where the spectrum is purely
real or purely imaginary and the spectrum under periodic boundary conditions fails to enclose a finite spectral area, a condition
typically associated with NHSE~\cite{salib2023arxiv,Okuma2020PRL,ZhangYangFang2022NatCommun,RiveroFengGe2022PRL,Hu2025Topological}.
A central challenge is therefore to determine which NH features leave unambiguous fingerprints in bulk observables accessible to transport and optical probes~\cite{Bergholtz2021RMP,Ashida2021AdvPhys}.
 In this paper, we show that a response-based
diagnostic of effective non-Hermiticity can be formulated using minimal pseudo-Lorentz-symmetry-breaking deformations,
which separate observables that remain captured by parameter redefinitions from those that exhibit irreducible NH
structure.

Non-Hermiticity makes the problem subtle because it can enter observables through two distinct mechanisms. In one route, it acts at the level of the spectrum, producing complex energies, exceptional degeneracies, and the accompanying nonanalytic features~\cite{Heiss2012ThePhysicsofEPs,Kawabata2019Classification}. In the other, the spectrum may remain entirely real, as in the weak-NH regime, yet the eigenstates are nonetheless altered, so that response functions probe the underlying biorthogonal wave-function structure~\cite{Brody2014Biorthogonal,Kunst2018Biorthogonal}. Crucially, these two effects do not have to track each other in bulk measurements: depending on whether a probe is controlled primarily by spectral properties or by the wave-function structure, it can be highly sensitive to one manifestation of non-Hermiticity while remaining largely insensitive to the other.
\begin{table*}[t]
\centering
\begin{tabular}{|
p{3.0cm}|
p{2.2cm}|
>{\centering\arraybackslash}p{4.8cm}|
>{\centering\arraybackslash}p{4.8cm}|
p{1.8cm}|}
\hline
\textbf{Observable} & \textbf{Scaling} & \textbf{Tilt deformation} & \textbf{VAD} & \textbf{Where} \\
\hline
$\rho(E)$ & $\sim |E|$ & $\sim f_{\rm tilt}(\alpha,\beta)$ { (\cmark)} & $\sim (v_x^{\rm eff}v_y^{\rm eff})^{-1}$ { (\xmark)} & Sec.~II \\
\hline
$Q_{\mu\nu}(k)$  & $\sim k^{-2}$ & $=Q^{(0)}_{\mu\nu}$ { (\xmark)}  & $=Q_{\mu\nu}(v_x^{\rm eff},v_y^{\rm eff})$ { (\xmark)} & Sec.~III \\
\hline
$\sigma^{(1)}_{ij}(\omega)$ & $\sim \omega^{0}$ &
$\sigma^{(1)}_{xx}=\sigma^{(1)}_{yy}=\sigma_0$ { (\xmark)} &
\begin{tabular}[c]{@{}c@{}}
$\sigma^{(1)}_{xx}=(v_x^{\rm eff}/v_y^{\rm eff})\sigma_0$\\
$\sigma^{(1)}_{yy}=(v_y^{\rm eff}/v_x^{\rm eff})\sigma_0$
\end{tabular} { (\xmark)}
& Sec.~IV\,A \\
\hline
$\sigma^{(2)}_{ijk}(\omega_1,\omega_2)$ & $\sim (\omega_1\omega_2)^{-1}$  & $\neq 0$ (inversion odd) { (\xmark)} & $=0$ (inversion even) { (\xmark)} & Sec.~IV\,B \\
\hline
$\eta_{\alpha\beta\gamma\delta}(\omega)$ & $\sim \omega^{2}$ & $=\eta_0(\omega)\,P_{\alpha\beta\gamma\delta}$ { (\xmark)} & inequivalent components { (\cmark)} & Sec.~V  \\
\hline
\end{tabular}
\caption{Behavior of the observables in a deformed two-dimensional ($d=2$) non-Hermitian Dirac semimetal  at zero temperature ($T=0$) with the dynamical exponent $z=1$: scaling exponents of observables are fixed by $(d,z)$ while  diagnostic content is in prefactors and tensor structure. Non-Hermiticity-insensitive benchmarks{, labeled  with a cross (\xmark),} are given by quantum geometric tensor [$Q_{\mu\nu}({\mathbf k})$ with $Q_{\mu\nu}^{(0)}({\mathbf k})$ being the form for the undeformed Dirac Hamiltonian, Eq.~\eqref{eq:NH-DSM-QGT}],   and linear optical conductivity  ($\sigma^{(1)}(\omega)$), with $\sigma_0$ as its universal value (Eq.~\eqref{eq:OC-sigma0}). Non-Hermiticity-sensitive  fingerprints{, labeled with a checkmark (\cmark),} are slope of the density of states ($\rho(E)$) for tilt deformation and  viscosity tensor [$\eta_{\alpha\beta\gamma\delta}(\omega)$] for velocity-anisotropic deformation (VAD). See Eqs.~\eqref{eq:viscosity-symm} and~\eqref{eq:eta_AS_projector}.  The last column indicates the section in which each observable is discussed.}
\label{tab:scaling_oneline_vlines}
\end{table*}

This challenge is especially pertinent at charge neutrality, where symmetry and the Dirac band structure often enforce
universal response coefficients. In particular, in minimal weak-NH Dirac models with real spectra several bulk
responses remain effectively NH insensitive: non-Hermiticity can be absorbed into a renormalization of band parameters
(e.g., an effective Dirac velocity), so that standard observables fail to provide independent access to NH
couplings~\cite{jurivcic2024yukawa}. Put differently, response functions can mask the microscopic origin of
non-Hermiticity, yielding results indistinguishable from those of a purely Hermitian theory with a suitably
reparametrized band structure. A guiding principle in this work is scaling: in the collisionless
regime at charge neutrality, relevant for Dirac semimetals, the dynamic exponent $z=1$ and spatial dimensionality $d=2$ fix the leading
power-law behavior of the responses, so the central issue is not the exponent itself but whether non-Hermiticity leaves
irreducible, response-dependent prefactors or tensor structures that survive any reparametrization. This motivates the search for   minimal, symmetry-allowed
perturbations and a set of experimentally accessible
observables for which NH parameters enter bulk response
in a way that cannot be reduced to such parameter
redefinitions.

\subsection{Summary of the results}

Here we show that minimal Hamiltonian deformations provide precisely such a controlled setting by sharply distinguishing responses that collapse under parameter redefinitions from those that retain
irreducible NH structure, as also summarized in Table~\ref{tab:scaling_oneline_vlines}. In Dirac and Weyl materials, the simplest departures from a perfectly pseudo-Lorentz-invariant cone, most
notably a {tilt} term~\cite{Goerbig2008Tilted,Trescher2015Quantum,Varykhalov2017Tilted,volovik2017lifshitz,Sikkenk2017Disorder,reiser2024tilted,Hu2025QC} and {velocity-anisotropy deformation} (VAD)~\cite{Moon2011Low-velocity,liu2014discovery,liu2014stable,Rodionov2015Effects,Mastrogiuseppe2016Hybridization,Pozo2018Anisotropic,Ma2025Prediction}, are ubiquitous already in Hermitian
systems. Their impact is apparent already at the level of basic spectral features. In both cases the density of states
remains linear, $\rho(E)\sim |E|$ but the {origin} of the slope is different: a tilt deformation can imprint an effective non-Hermiticity in the prefactor, whereas the VAD implements a controlled anisotropic deformation of the Dirac cone, equivalently introducing distinct
direction-dependent effective velocities $v_x^{\rm eff}$ and $v_y^{\rm eff}$. As a result, the DOS for VAD is insensitive to non-Hermiticity since it is modified through an anisotropic velocity rescaling.

A natural candidate for a complementary, wave-function-sensitive diagnostic of non-Hermiticity is  the biorthogonal quantum geometric tensor (QGT),
which extends the quantum metric and Berry curvature to non-Hermitian band structures~\cite{provost1980,Shen2018Topological,Ozawa2018SSHall,Zhang2019QGTinPTQM,Zhang2019NHquantumsystems,Zhu2021Band,Solnyshkov2021Quantum,Hu2024Generalized,Ozawa2025Geometric}. However,  as we here show, in the weak-NH, real-spectrum regime,
the quantum metric is insensitive to non-Hermiticity for both deformations considered here. For {tilt}, the
eigenstate geometry is unchanged, so the QGT remains {identical} to its untilted form. {Velocity anisotropy}
is likewise QGT-insensitive to non-Hermiticity in this regime: it modifies the QGT in the way already present in the Hermitian problem, without any additional dependence on the non-Hermitian
coupling beyond what can be absorbed into  effective anisotropic velocities. This stands in sharp contrast to the DOS, whose linear form is universal but its  {slope} can already acquire a
genuine NH dependence for the tilt deformation.

With this structure in place, we turn to bulk response functions. In Sec.~\ref{sec:optics} we compute the collisionless linear and second-order optical conductivities and show that the linear response is NH-insensitive: for the tilt it remains universal, while for the VAD it reduces to the standard anisotropic-Dirac form through a simple reparametrization in terms of effective velocities. The second-order conductivity is symmetry-selected, vanishing unless inversion symmetry is broken, and is therefore nonzero only for the tilt deformation (breaking this symmetry). We then consider stress-response observables in Sec.~\ref{sec:viscosity}, where the optical shear viscosity offers a
complementary window on non-Hermiticity through its tensorial structure. In particular, we find that the shear viscosity is anisotropic with multiple independent components for the velocity-anisotropy deformation (VAD), whereas it
collapses to an isotropic projector form for the tilted NH Dirac Hamiltonian. This identifies the DOS and viscoelastic response as natural bulk diagnostics, where minimal deformations  reveal effective non-Hermiticity.

\subsection{Organization}

Our paper is organized as follows. In Sec.~\ref{sec:model} we introduce the  NH Dirac model with minimal deformations  realizing the tilt and the VAD, and discuss   
the DOS for both deformations. In Sec.~\ref{sec:qgt} we analyze the biorthogonal QGT as a possible NH diagnostic. In Sec.~\ref{sec:optics} we turn to optical response and compute
the collisionless linear and second-order conductivities where we show that both are NH-insensitive. In Sec.~\ref{sec:viscosity} we study the optical shear viscosity and
show that stress response provides a complementary probe of non-Hermiticity: it is anisotropic with multiple
independent components for VAD, but collapses to an isotropic projector form for tilted NH Dirac Hamiltonian.
Finally, Sec.~\ref{sec:discussion} summarizes our results and discusses natural extensions. Technical details are presented  in the Appendixes.

\section{Model}
\label{sec:model}

We consider a minimal continuum description of a two-dimensional Dirac semimetal at charge neutrality in the
presence of weak non-Hermiticity. The low-energy degrees of freedom are collected into a four-component spinor
$\Psi_{\bm k}$, which in microscopic realizations can be viewed as carrying sublattice and valley flavors for
spinless Dirac materials (e.g., graphene)~\cite{neto2009}. Throughout we set $\hbar=1$ and work at $T=0$ unless
stated otherwise.

The Hermitian Dirac Hamiltonian reads
$\mathcal{H}_D=\sum_{\mathbf{k}}\Psi_{\mathbf{k}}^\dagger H_D(\bm k)\Psi_{\mathbf{k}}$, with
\begin{equation}
\label{eq:Hermitian-Ham}
H_D(\bm k)=v_{\rm H}\,h_0(\bm k),
\end{equation}
where the operator 
\begin{equation}
h_0(\bm k)=\Gamma_x k_x+\Gamma_y k_y,
\end{equation}
and the Dirac matrices obey the Clifford algebra
$\{\Gamma_i,\Gamma_j\}=2\delta_{ij}$ ($i,j=x,y$). The spinor $\Psi_{\mathbf{k}}$ is model dependent; for definiteness we
take it to be four-component, as in graphene with two sublattice and two valley degrees of freedom. We choose the
Hermitian representation $\Gamma_x=\sigma_1\otimes\tau_3$ and $\Gamma_y=\sigma_2\otimes\tau_0$, where
$(\sigma_0,\bm\sigma)$ [$(\tau_0,\bm\tau)$] act in sublattice (valley) space. All final results depend only on the
Clifford algebra and are therefore representation independent.

A minimal NH extension is generated by a Hermitian matrix $M$ that anticommutes with $h_0$,
$\{M,h_0\}=0$, so that $M h_0$ is anti-Hermitian~\cite{jurivcic2024yukawa}. Introducing the dimensionless NH parameter
$\beta\equiv v_{\rm NH}/v_{\rm H}\in\mathbb{R}$, we write
\begin{equation}
\label{eq:HNH0}
H_{\rm NH}(\bm k)=v_{\rm H}\,(1+\beta M)\,h_0(\bm k).
\end{equation}
The spectrum is
\begin{equation}
E_{\rm NH}(\bm k)=\pm v_{\rm F}|\bm k|,
\end{equation}
with the effective Fermi velocity 
\begin{equation}
v_{\rm F}=v_{\rm H}\sqrt{1-\beta^2},
\end{equation}
which is purely real for $|\beta|<1$. We focus on this weak-NH regime, where non-Hermiticity primarily enters
through the biorthogonal structure of eigenstates. 
For concreteness, we here choose $M=\sigma_3\otimes\tau_3$ to ensure spatial inversion symmetry of $H_{\rm NH}$, represented by
$I=\sigma_1\otimes\tau_1$, featured by its Hermitian counterpart in Eq.~\eqref{eq:Hermitian-Ham}. Importantly, many standard bulk responses at charge neutrality can then be
mapped to Hermitian forms by the replacement $v_{\rm H}\to v_{\rm F}$~\cite{jurivcic2024yukawa} which hinders the detection of non-Hermiticity; one of our goals is to identify minimal deformations and observables that can evade this limitation. 

\subsection{Minimal deformation terms: Tilt and velocity anisotropy}

To break pseudo-Lorentz symmetry without opening a gap, we add a (real) deformation term linear in $k_x$,
\begin{align}
\label{eq:Htilt}
H_{\rm NH}^{\rm def}(\bm k)
&=v_{\rm H}\Big[(1+\beta M)\,h_0(\bm k)+\alpha\,T\,k_x\Big]\nonumber\\
&\equiv H_{\rm NH}(\bm k)+H_{\rm def}(\bm k),
\end{align}
where $\alpha$ is a dimensionless parameter. Crucially, the character of the deformation is encoded in the commutation relation between the deformation matrix $T$
and the NH Dirac Hamiltonian. A tilt corresponds to a commuting deformation,
$[T,H_{\rm NH}(\bm k)]=0$, which can be realized, for example, by choosing $T=\sigma_0\otimes\tau_0$. In this case the
tilt acts as a momentum-dependent chemical potential, leading to the spectrum
\begin{equation}
E_{\rm NH}^{\rm tilt}(\bm k)=\pm v_{\rm F}|\bm k|+\alpha v_{\rm H}k_x,
\end{equation}
and throughout we restrict to the type-I regime $\alpha^2+\beta^2<1$, where charge neutrality coincides with a nodal
point. This deformation breaks spatial inversion symmetry of $H_{\rm NH}$ in Eq.~\eqref{eq:HNH0}, but   the symmetry of the tilt deformation plays no role in our analysis except for the
second-order optical conductivity (Sec.~\ref{sec:optics-nonlinear}).

On the other hand, the VAD is defined by the matrix anticommuting with NH Hamiltonian, $\{T,H_{\rm NH}(\bm k)\}=0$, with the spectrum effectively becoming anisotropic close to each Dirac cone, 
 \begin{equation}\label{eq:VAD-spectrum}
E_{\rm NH}^{\rm VAD}(\bm k)=\pm \sqrt{(v_x^{\rm eff})^2 k_x^2+(v_y^{\rm eff})^2k_y^2},
 \end{equation}
where the effective velocity in the $x-$($y-$)direction is $v_x^{\rm eff}=v_{\rm H}\sqrt{1+\alpha^2-\beta^2}$ ($v_y^{\rm eff}=v_{\rm H}\sqrt{1-\beta^2}$), and we consider  the regime $|\beta|<1$ ensuring the purely real spectrum of the deformed NH Hamiltonian. With the stage being set, we now turn to the analysis of the DOS.

\subsection{Density of states}

We compute the density of states (DOS) from the retarded Green's function,
\begin{equation}
\label{eq:DOS_def}
\rho(E)= -\frac{1}{\pi}\,{\rm Im} \int_{\bm k}\Tr\,G^{{\rm def},R}(E,\bm k),
\end{equation}
with $\int_{\bm k}\equiv\int\frac{d^2\bm k}{(2\pi)^2}$
and  $G^{{\rm def},R}(E,\bm k)$ obtained by analytic continuation $i\omega\to E+i\eta$, in the limit $\eta\to0^+$, of the Matsubara Green's function corresponding to the deformed NH Dirac Hamiltonian in Eq.~\eqref{eq:Htilt}, 
given by 
\begin{equation} \label{eq:Gtilt_def} G^{\rm def}(i\omega,\vb k)\equiv\big[i\omega-H_{\rm NH}^{\rm def}(\vb k)\big]^{-1}, 
\end{equation}
which implies the universal scaling form of the DOS, $\rho(E)\sim |E|^{\frac{d}{z}-1}f_\rho$, where $f_\rho$ is a dimensionless scaling  function of the dimensionless variables that depend on the form of the Hamiltonian. Therefore, in $d=2$ and for the Dirac system featuring the linear energy-momentum dispersion, implying $z=1$, this scaling form reduces to $\rho(E)\sim |E|f_\rho$. 

We first consider the tilt deformation yielding the Green's function in the form   
\begin{align} \label{eq:GF-tilt-compact} &G^{\rm tilt}(i\omega,\vb k) = \Big[ (i\omega+H_{\rm NH})\,A_{+} -B {\hat v}_M\,(\bm{\Gamma}\!\cdot\!\vb k)\,T\,k_x \nonumber\\ &- 2\alpha \vf k_x\Big( \vf^{2}k^{2}\,T +\alpha \vh k_x {\hat v}_M (\bm{\Gamma}\!\cdot\!\vb k) \Big) \Big]\, \frac{1}{B^{2}-A_-^{2}}. 
\end{align} 
Here,  $H_{\rm NH}$ is given in Eq.~\eqref{eq:HNH0}, ${\hat v}_M\equiv\vh+\vnh M$, and 
\begin{equation} \label{eq:ApmB_defs} A_{\pm}=\omega^{2}+\vf^{2}k^{2}\pm \alpha^{2}\vh^{2}k_x^{2}, \qquad B=2i\alpha \vh\omega k_x, 
\end{equation} 
with $k\equiv|\vb k|$ and $\vf=\vh\sqrt{1-\beta^2}$.

At charge neutrality the DOS remains linear, $\rho(E)\sim |E|$, as expected for a subcritical Dirac cone, with the explicit form~\cite{Pino-Alarcon2025Yukawa}
\begin{equation}
\label{eq:DOS_short}
\rho_{\rm tilt}(E)=N_f\frac{|E|}{\pi (v_{\rm F}^{\rm tilt})^2}\sqrt{\frac{1-\beta^2}{1-\alpha^2-\beta^2}},
\end{equation}
with $v_{\rm F}^{\rm tilt}=v_H\sqrt{1-\alpha^2-\beta^2}$, and  $N_f$ is the number of four-component fermion species. 
Equation~\eqref{eq:DOS_short} already shows that 
the DOS is consistent with the expected scaling form  but its  slope depends on $\alpha$ and $\beta$ in a way that is not captured by $\vf^{\rm tilt}$ alone, therefore representing an NH sensitive observable. Finally, in the limit $\alpha\to0$, the DOS reduces to the form for the deformation-free Hamiltonian with the NH parameter absorbed in the effective Fermi velocity. 

For the VAD, the DOS provides a simple illustration of NH
insensitivity in the real-spectrum regime. The deformation breaks isotropy by assigning different velocities along the two directions (see
Eq.~\eqref{eq:VAD-spectrum}), with the Green's function of the form,
\begin{equation}
G^{\rm VAD}(i\omega,\vb k) = - \frac{i\omega+H_{\rm NH}^{\rm VAD}}{\omega^2+(E_{\rm NH}^{\rm VAD}(\bm k))^2},
\end{equation}
where $H_{\rm NH}^{\rm VAD}$ is given by Eq.~\eqref{eq:Htilt} with the matrix $T$ anticommuting with the NH Dirac Hamiltonian [Eq.~\eqref{eq:HNH0}], yielding the DOS 
\begin{equation}\label{eq:DOS-VAD}
\rho(E)=N_f\frac{|E|}{\pi\,v_x^{\rm eff}v_y^{\rm eff}}. 
\end{equation}
Since the electronic bands are still described by an anisotropic Dirac cone, the DOS preserves its universal
linear scaling, $\rho(E)\sim |E|$.
 Crucially, any dependence on the NH coupling can be absorbed into a
redefinition of direction-dependent {effective} velocities, $v_x^{\rm eff}$ and $v_y^{\rm eff}$, so that the
resulting expression is identical to the DOS of an ordinary Hermitian anisotropic Dirac Hamiltonian. In this sense, VAD reshapes the DOS 
prefactor purely kinematically, by changing the effective velocities, without generating a
distinct fingerprint of non-Hermiticity in this observable.

\section{Quantum geometric tensor}
\label{sec:qgt}

While spectral diagnostics such as the DOS are fixed by the single-particle spectrum, they do not access the geometric information carried by the eigenvectors, i.e., the Hilbert-space embedding of the bands. We therefore begin with a wave-function-sensitive quantity: the biorthogonal quantum
geometric tensor (QGT), which generalizes the quantum metric and Berry curvature to NH band
structures~\cite{provost1980,berry1984,garrison1988}. In the weak-NH, real-spectrum regime considered here, the QGT
directly diagnoses whether a given tilt realization reshapes the left/right eigenstates and thus whether
non-Hermiticity leaves an irreducible imprint beyond parameter renormalizations. Importantly, the QGT is gauge invariant under smooth, momentum-dependent U(1) transformations of the biorthogonal eigenstates
that preserve the chosen normalization, and it can be expressed entirely in terms of the associated band projector.
It therefore provides a sharp, representation-independent notion of geometry of the 
Hilbert space for NH bands, even though the right/left eigenvectors need not be related by
any unitary transformation.

For band $n$ we define the QGT as 
\begin{equation}
\label{eq:QGT_def}
Q^n_{\mu\nu}=
\langle \partial_\mu \psi_n^L | \partial_\nu \psi_n^R \rangle
-
\langle \partial_\mu \psi_n^L | \psi_n^R \rangle
\langle \psi_n^L | \partial_\nu \psi_n^R \rangle,
\end{equation}
where $|\psi_n^{R/L}\rangle$ are right/left eigenstates of $H_{\rm NH}^{\rm tilt}(\bm k)$ and
$\partial_\mu\equiv\partial/\partial k_\mu$ with $\mu,\nu\in\{x,y\}$. The real symmetric part,
$g^n_{\mu\nu}\equiv \re\,Q^n_{\mu\nu}$, defines the (biorthogonal) quantum metric, which controls the infinitesimal
distance between nearby Bloch states in momentum space, while the imaginary antisymmetric part,
$\Omega^n_{\mu\nu}\equiv -2\,\im\,Q^n_{\mu\nu}$, plays the role of a (biorthogonal) Berry curvature. In the present
work we focus on the metric sector, since it provides the most direct diagnostic of whether tilt reshapes the
eigenstates in a way that is invisible to eigenvalue-only quantities.

In the untilted case the QGT is rotationally invariant and, in Cartesian coordinates, takes the form
\begin{equation}
\label{eq:NH-DSM-QGT}
\hat{Q}=
\frac{1}{4(k_x^2+k_y^2)^2}
\begin{pmatrix}
k_y^2 & -k_xk_y\\
-k_xk_y & k_x^2
\end{pmatrix},
\end{equation}
for both conduction and valence bands. Notably, Eq.~\eqref{eq:NH-DSM-QGT} is independent of velocity parameters, so
the replacement $v_{\rm H}\to v_{\rm F}$ leaves the QGT unchanged. Equivalently, upon transforming to polar
coordinates $(k,\varphi)$, the only nonzero component is $Q_{\varphi\varphi}=1/4$. This purely angular quantum geometry can be viewed as a direct imprint of the scale (and fixed-point conformal) structure
of the massless Dirac Hamiltonian, as follows. At charge neutrality the eigenprojectors depend only on the direction
$\hat{\bm k}=\bm k/|\bm k|$ and are invariant under the scale transformation $\bm k\to \lambda \bm k$, which rescales the
spectrum but leaves the eigenstates unchanged. Consequently, radial derivatives of the Bloch spinors vanish and the
quantum metric has no radial components, $Q_{kk}=Q_{k\varphi}=0$, while $Q_{\varphi\varphi}=1/4$ is fixed by the geometry of the Dirac cone as directly obtained from Eq.~\eqref{eq:NH-DSM-QGT}. In this sense the quantum geometry is entirely angular and independent of the radial component of momentum $k$, consistent with the
underlying scale invariance of the Dirac Hamiltonian.

Turning on the tilt, we find that the eigenstate geometry is unaffected: the tilt term acts as a purely spectral deformation that leaves the band projectors unchanged. Consequently, the right/left eigenvectors coincide with those of the untilted model and the QGT remains identical to Eq.~\eqref{eq:NH-DSM-QGT}, consistent with Ref.~\cite{ulrich-schnyder2025}. Put differently, tilt shifts the dispersion without rotating the pseudospin texture, so the Hilbert-space embedding of the bands is unmodified even though the single-particle energies are affected by the tilt term.

For the velocity-anisotropic Hamiltonian, the deformation reshapes the eigenstates, and the QGT takes the form
\begin{equation}
\label{eq:QGT_sym_tilt}
\hat{Q}_{\rm VAD}=
\frac{(1-\beta^2)(1-\beta^2+\alpha^2)}
{4\big[(1-\beta^2)k^2+\alpha^2k_x^2\big]^2}
\begin{pmatrix}
k_y^2 & -k_xk_y\\
-k_xk_y & k_x^2
\end{pmatrix}.
\end{equation}
That is, the matrix part retains the usual Dirac structure, while the prefactor encodes the deformation. In
particular, the prefactor makes the geometry explicitly anisotropic, with the degree of anisotropy controlled by $\alpha$, and the overall scale modulated by
the NH coupling $\beta$. This form (Eq.~\eqref{eq:QGT_sym_tilt}) reduces smoothly to the quantum metric in the undeformed case, given by Eq.~\eqref{eq:NH-DSM-QGT}, as $\alpha\to0$.
Moreover, for any finite deformation its $\beta$-dependence can be re-expressed as a rescaling of effective velocities
along the two directions (Eq.~\eqref{eq:VAD-spectrum}). In this sense, the resulting quantum metric is indistinguishable from
that of a {Hermitian} Dirac semimetal with anisotropic velocities. Thus, within the minimal real-spectrum Dirac models, the QGT does not furnish a direct bulk diagnostic of effective
non-Hermiticity: its NH dependence can be absorbed into velocity renormalizations rather than producing a
distinct geometric imprint.

\section{Optical conductivities}
\label{sec:optics}
We next turn to optical response in the collisionless regime at charge neutrality and  $T=0$, where interband
particle--hole excitations dominate, and the corresponding optical conductivities assume  universal forms. We compute the linear and second-order conductivities for {both} minimal deformations considered here, and show that they provide {NH-insensitive} benchmarks.
\begin{figure}[t!]
    \centering
    \includegraphics[scale=1]{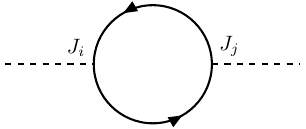}
    \caption{Polarization diagram corresponding to Eq.~\eqref{eq:Pij}. Here, a solid line denotes the Dirac fermion propagator, a dashed
    line corresponds to the external electromagnetic potential, while the vertex corresponds to the current as dictated by the U(1) gauge invariance.}
    \label{fig:Polarization}
\end{figure}

\subsection{Linear optical conductivity}
\label{sec:optics-linear}

We compute the collisionless linear optical conductivity using the Kubo formula.
Coupling to an external electromagnetic field via minimal substitution,
$\bm{k}\to \bm{k}-e\bm{A}$ (setting $\hbar=1$),
the Bloch Hamiltonian becomes $H(\bm{k}-e\bm{A})$. Expanding to linear order in
$\bm{A}$ defines the current operator
\begin{equation}
J_i(\bm{k})\equiv -\left.\frac{\partial H(\bm{k}-e\bm{A})}{\partial A_i}\right|_{\bm{A}=0}
= e\,\frac{\partial H(\bm{k})}{\partial k_i}
\equiv e\,v_i(\bm{k}),
\end{equation}
with $v_i$ the velocity operator and $i\in\{x,y\}$. The  linear conductivity is obtained from the  polarization (current-current correlator) at $T=0$, with the corresponding Feynman diagram in Fig.~\ref{fig:Polarization},  which reads as 
\begin{equation}
\Pi_{ij}(i\Omega)= -\int_{\nu, \bm{k}}
\mathrm{Tr}\left[
J_i(\bm{k})\,G(i\nu,\bm{k})\,
J_j(\bm{k})\,G(i\nu+i\Omega,\bm{k})
\right],
\label{eq:Pij}
\end{equation}
where $\int_{\nu,\bm{k}}\equiv \int \frac{d\nu d^2k}{(2\pi)^3}$ and
$G(i\omega,\bm{k})=[i\omega-H(\bm{k})]^{-1}$ is the (biorthogonal) single-particle Green's
function.  After analytic continuation $i\Omega\to \omega+i\eta$, with $\eta\to0^+$, the collisionless conductivity
follows from the retarded correlator as
\begin{equation}
\sigma_{ab}(\omega)=\frac{1}{\omega}\,{\rm Im}\,\Pi^{R}_{ab}(\omega),
\end{equation}
evaluated at charge neutrality and $T=0$, where only interband particle--hole processes contribute. The U(1) gauge invariance of the electromagnetic coupling then implies that the linear optical conductivity has the engineering scaling dimension ${\rm dim}[\sigma_{ab}]=d-2$, implying that it is dimensionless in $d=2$, and thus universal for the undeformed Dirac Hamiltonian, with the value  
\begin{equation}
\label{eq:OC-sigma0}
\sigma_0=N_f\frac{\pi}{4}
\end{equation}
in units $e^2/h$~\cite{SachdevQPT2011}.

For the tilt deformation the longitudinal optical conductivity retains its  universal form, which is
independent of $\alpha$ and $\beta$, as for  the Hermitian Dirac Hamiltonian, 
\begin{equation}
\label{eq:sigma_univ}
\sigma_{xx}(\omega)=\sigma_{yy}(\omega)=\sigma_0.
\end{equation}
Thus, in the collisionless interband
regime, the longitudinal linear optical conductivity provides an NH-insensitive response even in the presence of
tilt.

For the VAD, the longitudinal optical conductivity is no longer strictly universal,
but it remains {NH-insensitive} in the sense that it can be written in the
standard anisotropic-Dirac form, with non-Hermiticity entering only through the
effective velocities,
\begin{align}
\label{eq:sigma_VAD}
\sigma_{xx}(\omega)&= \sigma_0\,\mathcal{F}(\alpha,\beta),\nonumber\\
\sigma_{yy}(\omega)&= \sigma_0\,\mathcal{F}(\alpha,\beta)^{-1},
\end{align}
in units of $e^2/h$, where
\begin{equation}
\mathcal{F}(\alpha,\beta)=\sqrt{\frac{1+\alpha^2-\beta^2}{1-\beta^2}}
=\frac{v_x^{\rm eff}}{v_y^{\rm eff}}.
\end{equation}
Thus, unlike the tilt deformation, which preserves the universal value $\sigma_0$, the
VAD introduces anisotropy through an overall velocity ratio, with all dependence on the
anisotropy and NH parameters absorbed into the single combination $v_x^{\rm eff}/v_y^{\rm eff}$.
In this way, the VAD optical response maps directly onto that of a generic (Hermitian or NH)
anisotropic Dirac semimetal, for which anisotropy appears as
$\sigma_{xx}=\sigma_{0}\,v_y/v_x$ and $\sigma_{yy}=\sigma_{0}\,v_x/v_y$, and hence
$\sigma_{xx}\neq\sigma_{yy}$ unless $v_x=v_y$.

\begin{figure}[t!]
    \centering
    \includegraphics[scale=1]{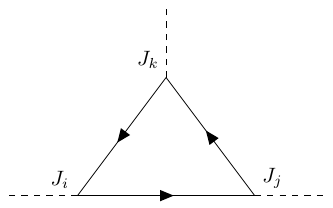}
    \caption{Second-order susceptibility (triangle) diagram corresponding to Eq.~\eqref{eq:chi2_def_main} yielding the second-order optical conductivity. The labels are the same as in Fig.~\ref{fig:Polarization}. }
    \label{fig:nlocfeynman}
\end{figure}

\subsection{Second-order optical conductivity}
\label{sec:optics-nonlinear}

For completeness, we next analyze the second-order optical response at charge neutrality and $T=0$. In the collisionless regime the
second-order conductivity tensor follows from the three-current Kubo formula~\cite{aversa1995,morimoto2016},
\begin{equation}
\label{eq:sigma2_def}
\sigma^{(2)}_{ijk}(\omega_1,\omega_2)=
-\lim_{\eta\to0^+}\sum_{\mathcal P}'\,
\frac{\chi^{(2)}_{ijk}(i\Omega_1,i\Omega_2)}{\omega_1\omega_2}
\Bigg|_{i\Omega_s\to \omega_s+i\eta},
\end{equation}
where $\sum_{\mathcal P}'$ symmetrizes under $(j,\omega_1)\leftrightarrow(k,\omega_2)$ and the corresponding 
susceptibility, with the Feynman diagram shown in Fig.~\ref{fig:nlocfeynman},
 reads
\begin{align}
\label{eq:chi2_def_main}
&\chi^{(2)}_{ijk}(i\Omega_1,i\Omega_2)
=\sum_{\mathcal P}\int_{\omega,\bm k}
\Tr\Big[
J_i(\bm k)\,G(i\omega,\bm k)\,
J_j(\bm k)\,\nonumber\\
&\times G(i\omega+i\Omega_1,\bm k)
J_k(\bm k)\,
G(i\omega+i\Omega_1+i\Omega_2,\bm k)
\Big].
\end{align}
This form of the susceptibility, together with the Kubo formula~\eqref{eq:sigma2_def}, implies the universal scaling form of the second-order optical conductivity for a two-dimensional ($d=2$) system with the dynamical exponent $z=1$, is $\sigma^{(2)}_{ijk}(\omega_1,\omega_2)=(\omega_1\omega_2)^{-1}f_{\sigma_2}$, where $f_{\sigma_2}$ is a universal dimensionless scaling function of the dimensionless variables~\cite{Rostami-PRR2020}. Furthermore, for this observable,  the spatial inversion symmetry, represented by the matrix $I=\sigma_1\otimes\tau_1$ for  the NH Dirac Hamiltonian in Eq.~\eqref{eq:HNH0} acts as a sharp symmetry diagnostic~\cite{MorimotoNagaosa2016}.  The tilt deformation, being odd under inversion allows a transverse nonlinear current,
\begin{align}
\label{eq:sigma2_as_final}
&\sigma^{(2)}_{{\rm tilt},yjk}(\omega_1,\omega_2)=
-\frac{i\,N_f e^3\,\alpha\,v_{\rm H}}{4\,\omega_1\omega_2}\;
\epsilon_{jk},
\end{align}
with $\epsilon_{jk}$ being completely antisymmetric symbol,  while the remaining components of the conductivity tensor are vanishing. See Appendix~\ref{sec:nloccalc} for technical details. 
On the other hand, for  the inversion-symmetric VAD, represented by $T=\sigma_3\otimes\tau_0$ for the choice of $M=\sigma_3\otimes\tau_3$ in the NH Dirac Hamiltonian~\eqref{eq:HNH0},
\begin{equation}
\label{eq:sigma2_sym_zero}
\sigma^{(2)}_{{\rm VAD},ijk}(\omega_1,\omega_2)=0.
\end{equation}
Therefore, for minimal deformations considered here,  the nonlinear
response is controlled by the  symmetry of the deformation term and its magnitude, while remaining independent of the NH strength
$\beta$, providing a second NH-insensitive optical response.

\section{Shear viscosity}
\label{sec:viscosity}

\begin{figure}[t!]
    \centering
    \includegraphics[scale=1]{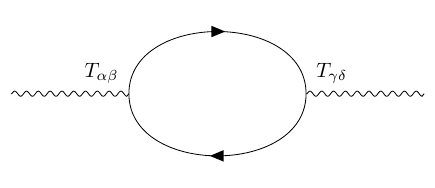}
    \caption{Feynman  diagram corresponding to the stress correlator in Eq.~\eqref{eq:Cstress_def} determining the optical shear viscosity.  Solid lines denote Dirac fermion propagators, wavy
    lines represent the external strain tensor, while the vertex corresponds to the stress tensor.}
    \label{fig:stress-correlator}
\end{figure}

We finally analyze the dynamical (optical) shear viscosity in the collisionless regime at $T=0$. Unlike the optical
conductivities in the minimal theory, the stress response is intrinsically sensitive to a reduction of  rotational
symmetry and to how the NH deformation reshapes the underlying biorthogonal structure, and it therefore provides a
natural setting in which deformation-dependent  NH effects may enter bulk response~\cite{tokatly2007,Bradlyn2012PRB,mueller2009,torre2015,EberleinPatelSachdev2017PRB,LinkSheehyNarozhnySchmalian2018PRB,LinkNarozhnyKiselevSchmalian2018PRL,CopettiLandsteiner2019PRB,Moore-PRB2020}. 

We define the stress tensor operator as~\cite{Bradlyn2012PRB}
\begin{equation}
\label{eq:stress_def}
T_{\alpha \beta}(\bm{k})=-i\,[H(\bm{k}),\mathcal{J}_{\alpha\beta}],
\end{equation}
where $\mathcal{J}_{\alpha\beta}=\mathcal{L}_{\alpha\beta}+S_{\alpha\beta}$ is the total angular momentum,
$\mathcal{L}_{\alpha\beta}=-x_{\alpha}k_{\beta}$ is orbital angular momentum, and $S_{\alpha\beta}$ generates
pseudospin rotations. The orbital part yields
\begin{equation}
\label{eq:stress_orb}
T_{\alpha \beta}^{(\rm o)}(\bm{k})=k_{\beta}\,\pdv{H(\bm{k})}{k_{\alpha}}.
\end{equation}
In the pseudospin sector, one may take $S_{ij}=-(1/4)\epsilon_{ijz}\sigma_z$. The (Matsubara) stress correlator is
\begin{align}
\label{eq:Cstress_def}
C_{\alpha \beta \gamma \delta}(i\Omega)=&
-\int_{-\infty}^\infty \frac{d \omega}{2 \pi}
\int \frac{d^2 \bm{k}}{(2\pi)^2}
\Tr\Big[
G(i \omega ,\bm{k})\, T_{\alpha \beta}(\bm{k})\nonumber\\
&\qquad\times
G( i \omega +i\Omega,\bm{k})\, T_{\gamma \delta}(\bm{k})
\Big],
\end{align}
with the Feynman diagram shown in Fig.~\ref{fig:stress-correlator}, and the optical viscosity follows from analytic continuation,
\begin{equation}
\label{eq:eta_def}
\eta_{\alpha \beta \gamma \delta}(\Omega)=\lim_{\delta\to0^+}
\frac{1}{\Omega}\,
{\rm Im}\Big[C_{\alpha \beta \gamma \delta}(i\Omega\to \Omega+i\delta)\Big],
\end{equation}
which, together with Eq.~\eqref{eq:Cstress_def}, implies the engineering scaling dimension of the shear viscosity ${\rm dim}[\eta_{\alpha \beta \gamma \delta}]=d$, and therefore in $d=2$ for a theory featuring dynamical exponent $z=1$, $\eta_{\alpha \beta \gamma \delta}(\Omega)=\Omega^2 {\mathcal G}_{\alpha \beta \gamma \delta}$, with $ {\mathcal G}_{\alpha \beta \gamma \delta}$ as the dimensionless function of the dimensionless variables which is determined by the details of the Hamiltonian.  

\begin{figure}[t!]
    \includegraphics[scale=0.45]{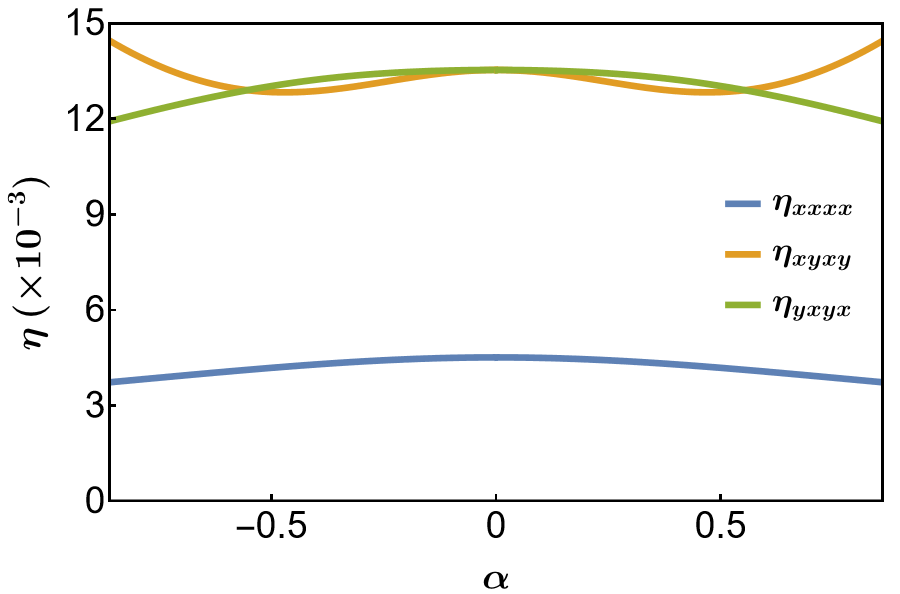}
    \caption{Components of the shear viscosity tensor for the velocity-anisotropy deformation with varying $\alpha$ at fixed $\beta=0.5$ at frequency $\Omega=1$ in the weak non-Hermitian regime with purely real spectrum given by Eq.~\eqref{eq:VAD-spectrum}.  We here fix $v_H=1$ and $N_f=1$. { See also Fig.~\ref{fig:eta1} for the plots for other values of the parameter $\beta$ fixed. }}
    \label{fig:etasym123-alpha}
\end{figure}

Turning first to the VAD,  the
viscosity develops a genuinely anisotropic multi-component structure with explicit dependence on $\alpha$ and $\beta$,
\begin{align}
\eta_{{\rm V},\rho\rho\rho\rho}(\Omega)
&=-\eta_{{\rm V},\rho\rho\sigma\sigma}(\Omega)
=-\eta_{{\rm V},\sigma\rho\rho\sigma}(\Omega)
=\Omega^2 f_1(\alpha,\beta),\nonumber\\
\eta_{{\rm V},xyxy}(\Omega)&=\Omega^2 f_2(\alpha,\beta),\quad\eta_{{\rm V},yxyx}(\Omega) =\Omega^2 f_3(\alpha,\beta),
\label{eq:viscosity-symm}
\end{align}
where $\rho,\sigma\in\{x,y\}$ with $\sigma\neq\rho$, and explicit forms of functions $f_i$, are given in   
Appendix~\ref{sec:visc}. 
The  decomposition in Eq.~\eqref{eq:viscosity-symm} makes transparent that, for VAD, the shear
viscosity tensor contains several independent components already at the level of collisionless response. The VAD reduces rotational symmetry and
the resulting viscoelastic response is no longer characterized by a single scalar function, but by a set of
inequivalent tensor components with amplitudes depending  on both \(\alpha\) and \(\beta\).

This structure is illustrated in Fig.~\ref{fig:etasym123-alpha}, which shows representative components as a function of the VAD strength
\(\alpha\) at fixed NH parameter \(\beta\). The curves are even in \(\alpha\), as expected since the effective velocities  are also even in this parameter. Importantly, the difference between the shear components $\eta_{xyxy}$ and $\eta_{yxyx}$ provides a direct diagnostic
of broken rotational invariance in the stress response. In a rotationally invariant Dirac theory these components are constrained to coincide; in the presence of the VAD, they become inequivalent, reflecting the deformation-induced anisotropy of the underlying biorthogonal structure.

\begin{figure}[t!]
    \centering
    \includegraphics[scale=0.43]{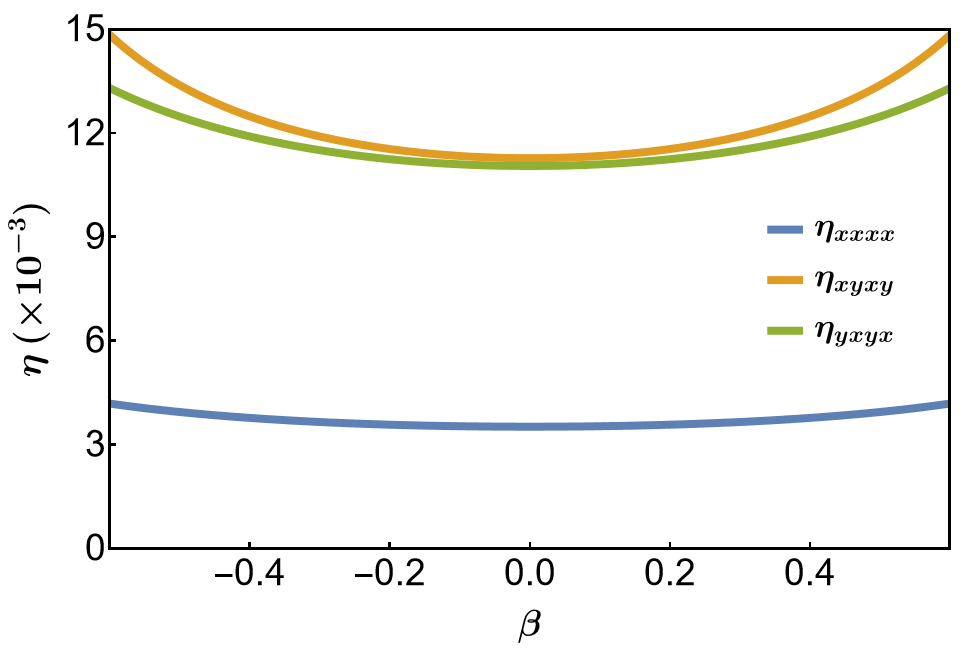}
    \caption{Components of the shear viscosity tensor for the velocity-anisotropy deformation with varying $\beta$ at fixed $\alpha=0.7$ at frequency $\Omega=1$ in the weak non-Hermitian regime with purely real spectrum given by Eq.~\eqref{eq:VAD-spectrum}. We here fix $v_H=1$ and $N_f=1$. { See also Fig.~\ref{fig:eta2} for the plots for other values of the parameter $\alpha$ fixed. }}
    \label{fig:etasym123-beta}
\end{figure}

For comparison, Fig.~\ref{fig:etasym123-beta} displays the same components as a function of the non-Hermiticity strength \(\beta\) at fixed \(\alpha\).
In this cut, the evolution with \(\beta\) isolates the
genuinely NH-sensitive part of the stress response: increasing \(|\beta|\) amplifies the anisotropic shear sector and
drives an overall enhancement that becomes strongest as one approaches the edge of the real-spectrum regime. Therefore,  for VAD, shear viscosity represents a  bulk probe of non-Hermiticity, which in the limit of vanishing deformation, $\alpha\to0$, reduces to the result for the NH Dirac Hamiltonian, given by  Eq.~\eqref{eq:eta_AS_scalar}. 

By contrast, for the  tilted NH Dirac Hamiltonian the response collapses to an isotropic form 
form,
\begin{equation}
\label{eq:eta_AS_projector}
\eta_{{\rm tilt},\alpha\beta\gamma\delta}(\Omega)=
\eta_{0}(\Omega)\,\mathcal{P}_{\alpha\beta\gamma\delta},
\end{equation}
with the projector
\begin{equation}
\label{eq:proj_def}
\mathcal{P}_{\alpha\beta\gamma\delta}=
\delta_{\alpha \gamma }\delta_{\beta \delta }+
\delta_{\alpha \delta}\delta_{\beta \gamma }-\frac{2}{d}\delta_{\alpha\beta}\delta_{\gamma\delta},
\end{equation}
and $d=2$. The scalar function $\eta_{0}(\Omega)$ can be written as
\begin{equation}
\label{eq:eta_AS_scalar}
\eta_{0}(\Omega)=\frac{N_f}{128}\left(\frac{\Omega}{v_{\rm H}\sqrt{1-\beta^2}}\right)^2,
\end{equation}
which coincides with the tilt-free Hermitian form upon the identification $v_{\rm F}=v_{\rm H}\sqrt{1-\beta^2}$.
Thus, together with the QGT, the shear viscosity separates NH-insensitive optical probes from stress-sensitive
observables that access effective non-Hermiticity in the real-spectrum weak-NH regime.

\section{Discussion and conclusions}
\label{sec:discussion}
A recurring difficulty in weakly non-Hermitian Dirac systems at charge neutrality is that many bulk responses can appear
``Hermitian-like,'' because the non-Hermitian coupling can often be absorbed into renormalized band parameters (most
notably an effective Dirac velocity), obscuring independent access to the non-Hermitian strength.
Our central message is that {minimal Hamiltonian deformations} act as a controlled filter: they separate observables fixed mainly by the dispersion from those that depend on how the eigenstates are deformed, and therefore reveal effective non-Hermiticity beyond a mere renormalization of band parameters. See also Table~\ref{tab:scaling_oneline_vlines}.

This separation is already apparent across the hierarchy of bulk probes. The DOS remains linear,
$\rho(E)\sim |E|$, but its slope distinguishes the two deformations: for type-I tilt it acquires an intrinsically
NH-dependent contribution that cannot be absorbed into a single effective velocity [Eq.~\eqref{eq:DOS_short}], whereas
for the VAD the result reduces to the standard anisotropic-Dirac form $\rho\sim |E|/(v_x^{\rm eff}v_y^{\rm eff})$
with all $\beta$-dependence absorbed into effective velocities [Eq.~\eqref{eq:DOS-VAD}]. A complementary wave-function
benchmark is provided by the biorthogonal QGT: the quantum metric is NH-insensitive for both deformations, being
unchanged for tilt [Eq.~\eqref{eq:NH-DSM-QGT}] and reducible to an anisotropic rescaling for VAD
[Eq.~\eqref{eq:QGT_sym_tilt}]. Consistently, the collisionless optical conductivities serve as robust NH-insensitive
benchmarks: the linear response is universal for tilt [Eq.~\eqref{eq:sigma_univ}] and maps onto the usual
anisotropic-Dirac form for VAD under $v_{x,y}\to v_{x,y}^{\rm eff}$ [Eq.~\eqref{eq:sigma_VAD}], in line with optical
sum rules linking interband optical weight to ground-state geometry
\cite{RestaSorella1999,SouzaWilkensMartin2000,VermaQueiroz2025Instantaneous,Resta2025Nonadiabatic}; at second order,
symmetry selects a nonzero signal only for inversion-breaking tilt
[Eqs.~\eqref{eq:sigma2_as_final},\eqref{eq:sigma2_sym_zero}]. By contrast, the shear viscosity sharply differentiates
the deformations: it collapses to an isotropic projector structure for tilt
[Eqs.~\eqref{eq:eta_AS_projector}--\eqref{eq:eta_AS_scalar}], but for VAD it develops inequivalent tensor components
with NH-dependent amplitudes [Eq.~\eqref{eq:viscosity-symm}], thereby singling out the DOS (tilt) and viscoelastic
response (VAD) as the observables where minimal deformations prevent NH effects from being fully absorbed into
parameter redefinitions.

A unifying outcome of our analysis is that, within the real-spectrum weak-NH regime, non-Hermiticity does not alter the
leading power-law scaling of the observables considered here, which is fixed solely by the Dirac dynamic exponent
$z=1$ and spatial dimensionality $d=2$. Instead, it enters through response-dependent amplitudes and tensor structures.
Depending on the probe, NH effects are either fully captured by effective parameter redefinitions or leave irreducible
signatures in terms of NH-dependent prefactors, anisotropies, or symmetry-selected components that provide a detectable measure of
effective non-Hermiticity.

{ On the other hand, the ``strong-NH'' regime $|\beta|>1$ is qualitatively different already at the level of the untilted
Dirac operator: the Dirac dispersion becomes purely imaginary, and once Hermitian deformations such as tilt are
included, the spectrum is generically complex. In this regime, the form of bulk response at real
frequencies depends on the underlying nonequilibrium steady state of the driven or dissipative system, including
gain/loss balance and noise. A quantitative treatment is therefore most naturally formulated within an open-system
framework, either using Keldysh nonequilibrium field theory or a Lindblad-type master equation, rather than
equilibrium Matsubara correlators~\cite{Sieberer2016RPP,Scarlatella2021PRX,Lindblad1976,Chaduteau2026PRL}. Nonetheless, one can still anticipate additional phenomena absent in the real-spectrum setting. In particular,
once a point gap opens the spectrum can develop exceptional structures, and wave-function- and stress-sensitive
observables may exhibit nonanalytic behavior across the exceptional threshold at $|\beta|=1$. A systematic analysis
of this regime, and of its consequences for optical and viscoelastic response, is left for future work.}

At present, the most experimentally available Dirac-type non-Hermitian platforms are topolectrical, photonic, and ultracold atom
realizations, where gain--loss and nonreciprocity can be engineered  with considerable flexibility and where the minimal deformations considered here (tilt and velocity anisotropy) can be implemented in a controlled manner~\cite{Zhang2024CLMcircuit,Zhu2023HigherRankTopolectrical,Xie2025ValleyLifetime,Yu2024DiracMassGainLoss,Zhen2015ExceptionalRing}.
A practical advantage of these platforms is that both the spectrum and spatial mode profiles are directly accessible, enabling straightforward measurements of DOS analogues and wave-function diagnostics even in regimes where standard bulk responses would otherwise appear effectively Hermitian-like. By contrast, in metamaterial implementations (topolectrical circuits and photonic lattices), the directly measured observables are
{effective linear-response functions of the engineered lattice model}, rather than electronic transport coefficients
such as charge optical conductivity or shear viscosity as in crystalline solids. In topolectrical circuits, the natural observable is the complex,
frequency-dependent {impedance matrix} (equivalently, the inverse circuit Laplacian), including
{nonlocal impedances} that reveal boundary resonances and bulk spectral features in close analogy to Green's-function
diagnostics \cite{Lee2018Topolectrical,Sahin2025TopolectricalReview}. In photonic platforms,
frequency-resolved absorption plays an analogous role; related protocols can reconstruct both real and imaginary parts
of \emph{non-Hermitian spectra} in momentum space \cite{Li2022NHAbsorption}. Ultracold atoms offer a complementary route in which non-Hermiticity is engineered via controlled dissipation: in a momentum-space lattice,
a dissipative Aharonov--Bohm chain has been realized that exhibits the NHSE, with Bragg spectroscopy resolving endpoint modes
\cite{Liang2022DynamicNHSEAtoms}. This approach is expected to provide a natural pathway toward two-dimensional NH Dirac platforms.
Beyond edge spectroscopy, closely related analogues of our bulk probes can be accessed through current  response
\cite{Anderson2019Conductivity}, circular-drive depletion (quantized dichroism) \cite{Asteria2019CircularDichroism}, and multi-tone
rectification protocols \cite{Lundh2005Ratchet}.

{ 
Since experimental systems are never perfectly clean, it is important to identify the regime in which the intrinsic collisionless response remains observable. In the weak-disorder regime, elastic scattering can be incorporated phenomenologically through a finite
broadening $\Gamma$ (or $1/\tau$), which rounds interband thresholds and smoothens  sharp features but does not alter the
symmetry-based features related to the deformations considered here. Our clean-limit results are therefore expected to
remain directly applicable for $\omega\gg \Gamma$, while the collision-dominated (hydrodynamic) regime, where $\omega\lesssim \Gamma$ and intraband (Drude/diffusive)
contributions thus become important, requires a separate treatment
beyond the scope of the present work.}

Several extensions follow naturally. A promising direction is to extend the analysis beyond the clean, charge-neutral, collisionless regime, and investigate how the
responses are modified once finite chemical potential and temperature, as well as disorder and interactions,
introduce intraband processes and finite broadening~\cite{Nissinen2017,KoziiFu2024PRB}. Likewise, entering the overtilted (type-II) regime should  markedly modify both optical and viscoelastic response structures. { Finally, a natural extension is to weakly NH Weyl semimetals in three dimensions, where their chiral Berry-monopole structure makes the corresponding linear and nonlinear responses a distinct problem left for future work.}
\section*{Acknowledgments}
We are grateful to Bitan Roy for the critical reading of the manuscript. This work was supported by Fondecyt (Chile) Grant No.~1230933 (V.J.). J.P.E. acknowledges support from Agencia Nacional de Investigaci\'on y Desarrollo (ANID), Doctorado Nacional, 2024-21240412.
\\


\appendix

\section{Nonlinear optical conductivity: evaluation of $\chi^{(2)}$}
\label{sec:nloccalc}

In this Appendix, we present the details of the calculation of nonlinear optical conductivity, which is related to the second-order susceptibility $ \chi_{i j k}^{(2)}(i\Omega_1,i\Omega_2)$, with the corresponding Feynman diagram in Fig.~\ref{fig:nlocfeynman}, where  
\begin{widetext}
\begin{equation}
       \chi_{i j k}^{(2)}(i\Omega_1,i\Omega_2) =\sum_{\mathcal{P}}\int\frac{d \omega}{2\pi}\int \frac{d^2 \vb{k}}{(2\pi)^2}\Tr\{J_i G_f(i \omega ,\vb{k})J_j   G_f(i \omega+i\Omega_1 ,\vb{k})J_k G_f(i \omega+i\Omega_1+i\Omega_2 ,\vb{k})\} 
\end{equation}
For this, we will show the details of the calculation of the  $ \chi_{yxy}^{(2)}(i\Omega_1,i\Omega_2)$ for the antisymmetric tilt case, showing that this method serves for all other components of the tensor. After completing the trace and integrating over the frequencies using residues, we obtain:
\begin{align}
& \chi_{yxy}^{(2)}(i\Omega_1,i\Omega_2)= 4 e^3 v_H^3\left(1-\beta ^2\right)^{3/2}\sum_{\mathcal{P}}\int \frac{dk d\varphi }{(2\pi)^2}   k  \cos (\phi )\Bigg(8 \left(\beta ^2-1\right)^2 k^4 v_H^4 (3 \cos (2 \phi )-1) \notag \\ &+8 i \alpha  \left(1-\beta ^2\right) k^3 \left(\Omega _1+2 \Omega _2\right) v_H^3 \cos (\phi )+2 \left(1-\beta ^2\right) k^2 v_H^2 \left(2 \Omega _1^2 \cos ^2(\phi )+\Omega _2 \Omega _1 (\cos (2 \phi )-3)+\Omega _2^2 (\cos (2 \phi )-3)\right)\notag\\&-2 i \alpha  k \Omega _1^2 \left(\Omega _1+2 \Omega _2\right) v_H \cos (\phi )-\Omega _1^2 \Omega _2 \left(\Omega _1+\Omega _2\right)\Bigg)\notag\\&\times \frac{1}{\left(2 \epsilon _k-i \Omega _1\right) \left(2 \epsilon _k+i \Omega _1\right) \left(2 \epsilon _k-i \Omega _2\right) \left(2 \epsilon _k+i \Omega _2\right) \left(2 \epsilon _k-i \left(\Omega _1+\Omega _2\right)\right) \left(2 \epsilon _k+i \left(\Omega _1+\Omega _2\right)\right)}
\end{align}
where \(\epsilon_k=\sqrt{1-\beta ^2} k v_H\). Now, we perform the analytical continuation $i\Omega_m\to \omega_m+i \delta $, and perform  the partial fraction decomposition of the last factor to find
   \begin{equation}
\int dk\, d\varphi\,
\frac{f(k,\omega_m)}{\omega_m-\epsilon_k+i\delta}
=
\int d k\, d\varphi\,
f(k,\omega_m)
\left[
\mathcal{P}\!\left(\frac{1}{\omega_m-\epsilon_k}\right)
- i\pi \delta(\omega_m-\epsilon_k)
\right].
\end{equation}
   \end{widetext}
   after carrying out  the analytical continuation. Subsequently, integration over the delta functions yields 
   \begin{equation}
     \chi_{yxy}^{(2)}(i\Omega_m\to \omega_m)=\sum_{\mathcal{P}}   -\frac{i  N_f e^3 \alpha \vh}{16  }
   \end{equation}
   Now, we carry out the sum over the permutation $\omega_1\to \omega_2$, and obtain the second-order linear conductivity:
\begin{equation}
 \sigma_{yxy}^{(2)}(\omega_1,\omega_2) =-\sum_{\mathcal{P}}'   \frac{\chi_{y x y}^{(2)}(i\Omega_1,i\Omega_2) }{\omega_1 \omega_2}\eval_{i\Omega_m\to \omega_m+i \delta}
 \end{equation}
 \begin{equation}
  \sigma_{yxy}^{(2)}(\omega_1,\omega_2)=   -\frac{i  N_f e^3 \alpha \vh}{8\omega_1\omega_2  }
\end{equation}
For the $xxx$ and $xyy$ components, we find 
\begin{equation}
     \chi_{xxx}^{(2)}(i\Omega_m\to \omega_m)=\chi_{xyy}^{(2)}=N_f\sum_{\mathcal{P}}  -\frac{i \alpha  e^3 \left(\omega _1-\omega _2\right) v_H}{4 \left(\omega _1+\omega _2\right)}
\end{equation}
When we sum over frequencies, due to its antisymmetry, both components vanish,
\begin{equation}
    \sigma_{xxx}^{(2)}(\omega_1,\omega_2)=\sigma_{xyy}^{(2)}(\omega_1,\omega_2)=0.
\end{equation}

\section{Viscosity tensor for the  velocity-anisotropy deformation}
\label{sec:visc}

We here show the explicit expressions for the three independent dimensionless  scaling  functions $f_{1,2,3}(\alpha,\beta,\Omega)$ for the components of the viscosity tensor in the case of  the VAD, as given  by the scaling form in  Eq.~\eqref{eq:viscosity-symm}. They  
can be expressed  in terms of complete elliptic integrals $K(m)$ and $E(m)$ as follows: 

\begin{widetext}
\begin{align}
&f_1(\alpha,\beta,\Omega)=\frac{N_f}{48 \pi  \alpha ^4 v_H}\Bigg( \sqrt{1-\beta ^2} \left(\alpha ^2-2 \beta ^2+2\right) E\left(\frac{\alpha ^2}{\beta ^2-1}\right)+\left(\alpha ^2-2 \beta ^2+2\right) \sqrt{\alpha ^2-\beta ^2+1} E\left(\frac{\alpha ^2}{\alpha ^2-\beta ^2+1}\right)\nonumber \\
&+2 \sqrt{1-\beta ^2} \left(-\alpha ^2+\beta ^2-1\right) K\left(\frac{\alpha ^2}{\beta ^2-1}\right)+2 \left(\beta ^2-1\right) \sqrt{\alpha ^2-\beta ^2+1} K\left(\frac{\alpha ^2}{\alpha ^2-\beta ^2+1}\right)\Bigg),\\
&f_2(\alpha,\beta,\Omega)
=\frac{N_f}{48\pi\,\alpha^{4}(\beta^{2}-1)\,(\alpha^{2}-\beta^{2}+1)^{3/2}\,v_H}
\Bigg[
\sqrt{(\beta^{2}-1)\,(-\alpha^{2}+\beta^{2}-1)}\;
E\!\left(\frac{\alpha^{2}}{\beta^{2}-1}\right)\nonumber\\
&\times\Bigl(8\alpha^{2}(\beta^{2}-1)^{2}-2(\beta^{2}-1)^{3}-2\alpha^{6}-\alpha^{4}(\beta^{2}-1)\Bigr)
\nonumber\\
&\qquad
-(\alpha^{2}-\beta^{2}+1)\Bigl(\alpha^{4}+7\alpha^{2}(\beta^{2}-1)-2(\beta^{2}-1)^{2}\Bigr)(\beta^{2}-1)\;
K\!\left(\frac{\alpha^{2}}{\alpha^{2}-\beta^{2}+1}\right)
\nonumber\\
&\qquad
-\sqrt{(\beta^{2}-1)\,(-\alpha^{2}+\beta^{2}-1)}\;
K\!\left(\frac{\alpha^{2}}{\beta^{2}-1}\right)
\nonumber\\
&\qquad
-(\alpha^{2}-\beta^{2}+1)\Bigl(2\alpha^{6}+\alpha^{4}(\beta^{2}-1)-8\alpha^{2}(\beta^{2}-1)^{2}
+2(\beta^{2}-1)^{3}\Bigr)\;
E\!\left(\frac{\alpha^{2}}{\alpha^{2}-\beta^{2}+1}\right)
\Bigg],\\
&f_3(\alpha,\beta,\Omega)
=\frac{N_f}{48\pi\,\alpha^{4}\sqrt{1-\beta^{2}}\bigl(-\alpha^{2}+\beta^{2}-1\bigr)\,v_H}
\Bigg[
3(\beta^{2}-1)\Bigl(3\alpha^{4}-2\alpha^{2}(\beta^{2}-1)-2(\beta^{2}-1)^{2}\Bigr)\,
E\!\left(\frac{\alpha^{2}}{\beta^{2}-1}\right)
\nonumber\\
&\qquad
-\alpha^{2}\bigl(3\alpha^{2}+\beta^{2}-1\bigr)\,
\sqrt{-\bigl(\alpha^{2}+2\bigr)\beta^{2}+\alpha^{2}+\beta^{4}+1}\;
K\!\left(\frac{\alpha^{2}}{\alpha^{2}-\beta^{2}+1}\right)
\nonumber\\
&\qquad
+\Bigl(6\alpha^{6}-13\alpha^{4}(\beta^{2}-1)+\alpha^{2}(\beta^{2}-1)^{2}
+6(\beta^{2}-1)^{3}\Bigr)\,
K\!\left(\frac{\alpha^{2}}{\beta^{2}-1}\right)
\nonumber\\
&\qquad
+\Bigl(3\alpha^{4}-2\alpha^{2}(\beta^{2}-1)-2(\beta^{2}-1)^{2}\Bigr)\,
{\rm Im}\Biggl\{
\alpha^{2} K\!\left(\frac{-\alpha^{2}+\beta^{2}-1}{\beta^{2}-1}\right)
-(\beta^{2}-1)\,E\!\left(\frac{-\alpha^{2}+\beta^{2}-1}{\beta^{2}-1}\right)
\Biggr\}
\Bigg].
\end{align}

{ We display the form of the  components of the shear viscosity tensor  as a function of the velocity-anisotropy parameter  $\alpha$ for   fixed values of the NH parameter $\beta$ in Fig.~\ref{fig:eta1}, while in Fig.~\ref{fig:eta2} we show the components of the same tensor as a function of $\beta$ for fixed values of the parameter $\alpha$. 

\begin{figure}[t!]
    \includegraphics[scale=0.237]{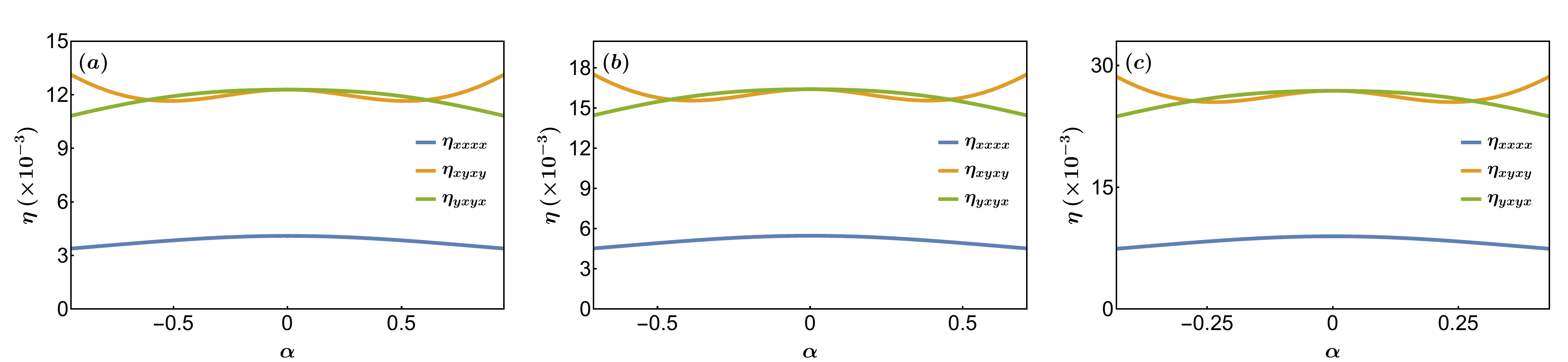}
    \caption{Components of the shear viscosity tensor for the velocity-anisotropy deformation with varying $\alpha$ at fixed (a) $\beta=0.3$, (b) $\beta=0.7$ and (c) $\beta=0.9$ and at frequency $\Omega=1$ in the weak non-Hermitian regime with purely real spectrum given by Eq.~\eqref{eq:VAD-spectrum}. We here fix $v_H=1$ and $N_f=1$. }
    \label{fig:eta1}
\end{figure}

\begin{figure}[t!]
    \includegraphics[scale=0.22]{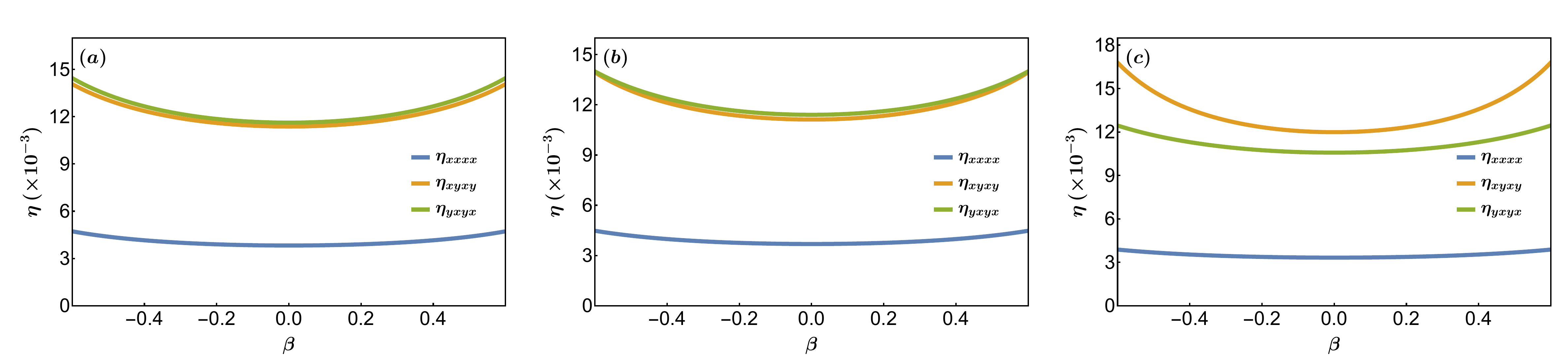}
    \caption{Components of the shear viscosity tensor for the velocity-anisotropy deformation  with varying $\beta$ at fixed (a) $\alpha=0.3$, (b) $\alpha=0.5$ and (c) $\alpha=0.9$, and at frequency $\Omega=1$ in the weak non-Hermitian regime with purely real spectrum given by Eq.~\eqref{eq:VAD-spectrum}. We here fix $v_H=1$ and $N_f=1$. }
    \label{fig:eta2}
\end{figure}}
\end{widetext}

\bibliography{references}

\end{document}